\documentclass[english,12pt]{article}
\usepackage{array}
\usepackage{graphicx}
\usepackage{amssymb}
\usepackage{amsmath}
\usepackage{multirow}
\usepackage{prettyref}
\usepackage{babel}
\usepackage{units}
\usepackage[latin1]{inputenc}
\usepackage{amsfonts}
\usepackage{amssymb}
\usepackage{babel}
\usepackage{cite}
\def\@fmsl@sh#1#2#3{\m@th\ooalign{$\hfil#1\mkern#2/\hfil$\crcr$#1#3$}}
 \def\eq#1\en{\begin{equation}#1\end{equation}}
\def\s[#1,#2]{[#1\stackrel{\star}{,}#2]}
\def\sx[#1,#2]{[#1\stackrel{\star_{x}}{,}#2]}

\newcommand{\nc}{\newcommand}
\nc{\beq}{\begin{equation}}
\nc{\eeq}{\end{equation}}
\nc{\beqa}{\begin{eqnarray}}
\nc{\eeqa}{\end{eqnarray}}

\def\bc{\begin{center}}
\def\ec{\end{center}}

\def\to{\rightarrow}

\def\gsim{\mathrel{\mathpalette\atversim>}}

\def\bc{\begin{center}}
\def\ec{\end{center}}

\def\gsim{\mathrel{\rlap{\lower4pt\hbox{\hskip1pt$\sim$}}

    \raise1pt\hbox{$>$}}}       

\def\gsim{\mathrel{\rlap{\lower4pt\hbox{\hskip1pt$\sim$}}
    \raise1pt\hbox{$>$}}}       

\begin{document}
\makeatletter
\def\fmslash{\@ifnextchar[{\fmsl@sh}{\fmsl@sh[0mu]}}
\def\fmsl@sh[#1]#2{%
  \mathchoice
    {\@fmsl@sh\displaystyle{#1}{#2}}%
    {\@fmsl@sh\textstyle{#1}{#2}}%
    {\@fmsl@sh\scriptstyle{#1}{#2}}%
    {\@fmsl@sh\scriptscriptstyle{#1}{#2}}}
\def\@fmsl@sh#1#2#3{\m@th\ooalign{$\hfil#1\mkern#2/\hfil$\crcr$#1#3$}}
\makeatother

\boldmath
\begin{center}
  \Large {\bf  Effective theory for quantum gravity}
    \end{center}
\unboldmath
\vspace{0.2cm}
\begin{center}
{  {\large Xavier Calmet}\footnote{x.calmet@sussex.ac.uk} }
 \end{center}
\begin{center}
{\sl Physics $\&$ Astronomy, 
University of Sussex,   Falmer, Brighton, BN1 9QH, UK 
}
\end{center}
\vspace{1cm}
\begin{center}
 Submission date: March 14, 2013 
\end{center}

\vspace{5cm}
\begin{abstract}
\noindent
In this paper, we discuss an effective theory for quantum gravity and discuss the bounds on the parameters of this effective action. In particular we show that measurement in pulsars binary systems are unlikely to improve the bounds on the coefficients of the $R^2$ and $R_{\mu\nu} R^{\mu\nu}$  terms obtained from probes of Newton's potential performed on Earth. Furthermore, we argue that if the coefficients of these terms are induced by quantum gravity, they should be at most of order unity since $R^2$ and $R_{\mu\nu} R^{\mu\nu}$ are dimension four operators. The same applies to the non-minimal coupling of the Higgs boson to the Ricci scalar.
\end{abstract} 
 \vspace{1cm}
Essay written for the Gravity Research Foundation 2013 Awards for Essays on Gravitation. 



\newpage

Finding a quantum formulation of general relativity  \cite{Einstein:1916vd} is notoriously difficult. Despite several interesting proposals see e.g.  \cite{Kiefer:2004gr} for a review, we are still far away from having a satisfactory quantum mechanical description of gravity. It would thus be helpful to have  some guidance from nature. Most experiments designed to date to test quantum gravitational effects rely on the hope that some basic symmetry of nature is violated by quantum gravitational effects or that the dispersion relation of light is modified. Here we will describe a framework which enables one to probe quantum gravitational effects directly without having to make speculative assumptions. The only hypothesis is that general covariance is the correct symmetry of gravity at the quantum level as well.

 When a fundamental theory is not known, the notion of effective field theory can be useful. Indeed it has proven to be extremely successful in different branches of physics ranging from particle physics to condensed matter physics. Effective theories are appropriate when one considers experiments at energies well below the scale at which the full underlying physics is expected to become apparent. Clearly this  is the case of quantum gravity. Physics experiments can be performed typically at  center of mass energies going up to 300 TeV if we think of high energetic cosmic rays, while quantum gravity is expected to become important at some energy scale $M_\star$ traditionally identified with the reduced Planck scale $ M_P=2.4335 \times 10^{18}$ GeV. The hierarchy between these two scales is the reason why it is so tough to probe quantum gravity experimentally.

While general relativity is very successful on macroscopic scales,  it is well known that its quantization is not straightforward. Indeed, general relativity  is not renormalizable, at least in perturbation theory.  But, assuming that we know the fundamental symmetry of nature, namely diffeomorphism invariance, and that gravity can be described by a massless spin 2 particle,  one can formulate an effective theory for quantum general relativity  \cite{Donoghue:1994dn,Donoghue:2012zc,Burgess:2003jk} valid up to the energy scale at which quantum gravitational effects become strong  $M_\star$. 

The effective field theory containing the metric $g_{\mu\nu}$, which describes the graviton if linearized, a cosmological constant  and the standard model of particle physics $\mathcal{L}_{SM}$ (Higgs doublet $H$ included) is given by
\begin{eqnarray}\label{action1}
S = \int d^4x \, \sqrt{-g} \left[ \left( \frac{1}{2}  M^2 + \xi H^\dagger H \right)  \mathcal{R}- \Lambda_C^4 + c_1 \mathcal{R}^2 + c_2 \mathcal{R}_{\mu\nu}\mathcal{R}^{\mu\nu}  + \mathcal{L}_{SM} + \mathcal{O}(M_\star^{-2})   \right] 
\end{eqnarray}
This effective action is an expansion in space-time curvature. The Higgs boson has a non-zero vacuum expectation value, $v=246$ GeV. The parameters $M$ and $\xi$ are then determined by 
\begin{eqnarray}
\label{effPlanck}(M^2+\xi v^2)=M_P^2 \, .
\end{eqnarray}
This effective field theory contains several energy scales. There is the reduced Planck scale $M_P$ of the order of $10^{18}$ GeV (equivalently Newton's constant),  the cosmological constant $\Lambda_C$ of order of $10^{-3}$ eV and the $M_\star$ which is traditionally identified with $M_P$ but this needs not to be the case if we do not trust this effective theory up to that energy scale. In the language of effective field theory, the coefficients $c_{1/2}$ and $\xi$ are called Wilson coefficients. 

As we know so little about quantum gravity,  it is  not clear how many of the Wilson coefficients of this effective action are fundamental parameters of nature, i.e. new coupling constants, or calculable using other parameters of the effective action. This distinction is important. For example, the Wilson coefficients of dimension four operators are expected to be very tiny if they are generated by some new physics at a scale $\Lambda_{NP}$. Indeed, in the limit where $\Lambda_{NP} \to \infty$ they must disappear. Hence they must be of the form $\exp{(-\lambda/\Lambda_{NP})}$ where $\lambda$ is identified with the ultra violet cutoff of the theory. Here we need to be more careful. When one expands $g_{\mu\nu}$ around a constant background metric $\eta_{\mu\nu}$ using $g_{\mu\nu}=\eta_{\mu\nu}+h_{\mu\nu}/M_P$, one finds that the dimension four operator $\xi H^\dagger H  \mathcal{R}$ is actually a dimension 6 operator of the type $\xi H^\dagger H h \Box h/M_P^2$ where $h$ is the graviton. The remaining dimension 4 operators of the type $\mathcal{R}^2$ are actually dimension 8 operators $h \Box h h \Box h/M_P^4$. One thus expects  that $\xi$, $c_1$ and $c_2$ are of order unity. On the other hand they might be arbitrarily large if they are genuinely new fundamental parameters of nature. The Wilson coefficient of higher order terms in $\mathcal{R}$, if induced by quantum gravity, are proportional to $(1/M_P)^n (\lambda/M_P)^m$ and thus expected to be small unless again they are fundamental parameters not calculable in terms of $M_P$.

One should stress that it is not clear what is the cutoff $\lambda$ for this effective theory, it might be $M_P$,  an energy scale corresponding to the cosmological constant scale or maybe even some other particle physics scale such as the weak scale or some other scale of grand unification. The smallness of the  observed cosmological constant could be the sign that the cutoff should be of the order of the cosmological constant scale.  However, the fact that we see no sign of new physics beyond usual general relativity at this energy scale could be interpreted  as failure of the naturalness argument just like the discovery of a single Higgs boson without any sign of new physics so far, seems to indicate that naturalness is not a helpful argument in our quest for physics beyond the standard model.

We shall now describe the current bounds on the values of the parameters of the quantum gravitational action given in Eq. (\ref{action1}) and discuss whether future experiments are likely to improve them or not. Stelle has shown \cite{Stelle:1977ry} that  the terms  $c_1 \mathcal{R}^2$ and $c_2 \mathcal{R}^{\mu\nu}\mathcal{R}_{\mu\nu}$ lead to Yukawa-like corrections to Newton's potential of a point mass $m$: 
\begin{eqnarray}
\Phi(r) = -\frac{Gm}{r} \left( 1+\frac{1}{3}e^{-m_0 r}-\frac{4}{3}e^{-m_2 r} \right) 
\end{eqnarray}
with
\begin{eqnarray}
m_0^{\,-1}=\sqrt{32\pi G \left(3c_1-c_2 \right)}
\end{eqnarray}
and
\begin{eqnarray}
m_2^{\,-1}=\sqrt{16\pi G c_2}.
\end{eqnarray}
Sub-millimeter tests of Newton's law \cite{Hoyle:2004cw} using sophisticated pendulums are used to bound $c_1$ and $c_2$. One finds that,  in the absence of accidental fine cancellations between both Yukawa terms, they are  constrained to be less than $10^{61}$ \cite{Calmet:2008tn} .

Astrophysical observations lead to bounds on these terms \cite{Psaltis:2008bb}. Binary systems of pulsars are the most promising environment to probe gravity at high curvature. However, a back of the envelop estimate quickly reveals that astrophysical observations will not be able to compete with Earth-based  tests of Newton's law. Let us approximate the Ricci scalar in the binary system of pulsars by $G M/(r^3c^2)$ where $M$ is the mass of the pulsar and $r$ is the distance to the center of the pulsar. Clearly, if the distance is larger than the radius of the pulsar, then the Ricci scalar vanishes, so we are doing a rather crude estimate. Let us be optimistic and assume we could probe gravity at the surface of the pulsar, we thus take $r=13.1$ km and M=2 solar masses. One then requests that the $R^2$ term should become comparable to the leading order Einstein-Hilbert term $\frac{1}{2}  M_P^2 R$ and finds that one could reach only bounds of the order of $10^{78}$ on $c_1$. Similar weak bounds are expected on $c_2$. Such limits are obviously much weaker that those obtained on Earth. 

We now consider the bounds on the non-minimal coupling of the Higgs boson to the Ricci scalar which is the third dimension four operator of the effective action.  The fact that the boson discovered at CERN behaves very much like what is expected of the Higgs boson of the standard model of particle physics, enables one to set a limit on the non-minimal coupling as for large $\xi$ the Higgs boson would decouple from the standard model. One finds $|\xi| >2.6 \times 10^{15}$ is excluded at the $95 \%$ C.L.  \cite{Atkins:2012yn}. Future colliders will not be able to improve this bound much and only potentially by one order of magnitude at a hypothetical future linear collider.

There is no bound to date on the coefficients of higher dimensional operators. Because they are suppressed by powers of the Planck scale, these terms are expected to be completely irrelevant unless their Wilson coefficients are unnaturally large.

Any progress towards measuring the parameters of the effective action will clearly require to be very creative. While this is clearly a difficult task, it  may be the only way to probe directly quantum gravity and to differentiate empirically  between different frameworks to quantize gravity. We stress that although the parameters of the quantum gravitational effective action are expected to be small if they are generated via quantum gravitational corrections, they could be large if they are truly independent parameters and not calculable in terms of the Planck mass. Indeed, who could have guessed without experimental guidance that gravity would be that much weaker that the weak interactions? The reason for this phenomenon is that the Planck mass is so large. Why wouldn't the other coefficients of the effective action be large as well? It is thus important to continue the ongoing experimental and observational efforts to measure these parameters. One should also keep in mind that one of the major theoretical development of the last 15 years has been the realization that $M_\star$, could be much below the traditional $2.4335 \times 10^{18}$ GeV if there are extra-dimensions with a large volume \cite{ArkaniHamed:1998rs,Randall:1999ee} or even in four space-time dimensions if there is a large hidden sector of particles \cite{Calmet:2008tn}. Clearly the LHC has been leading the way by setting some of the tightest limits to date on the experimental value of the Planck scale which we now know is above a few TeVs, but more creative ideas could lead to stronger bounds on this parameter of the quantum gravitational effective action.



\bigskip{}

\end{document}